\pdfoutput=1 
\documentclass[10pt,conference]{IEEEtran}
\AtBeginDocument{%
  \providecommand\BibTeX{{%
    \normalfont B\kern-0.5em{\scshape i\kern-0.25em b}\kern-0.8em\TeX}}}

\usepackage{mdframed}
\usepackage{graphicx}
\usepackage{templates/template}
\usepackage{multirow}
\usepackage{graphicx}
\usepackage{enumitem}
\usepackage{wrapfig}
\usepackage[capitalise]{cleveref}
\usepackage{array,multirow,graphicx}
\usepackage{hyperref}

\usepackage{xcolor}
\def\BibTeX{{\rm B\kern-.05em{\sc i\kern-.025em b}\kern-.08em
    T\kern-.1667em\lower.7ex\hbox{E}\kern-.125emX}}
    
\usepackage{amsmath}
\usepackage{booktabs} % For formal tables
\usepackage{subcaption}
\usepackage{colortbl}
\usepackage{color}
\usepackage{listings}
\usepackage{balance}
\usepackage{colortbl}
\usepackage{tcolorbox}
\usepackage{url}
\usepackage{soul}
\usepackage{tcolorbox}
\usepackage{MnSymbol}

\usepackage[font=small,labelfont=bf]{caption}
\usepackage{tcolorbox}
\definecolor{light-green}{rgb}{.5,1,.5}
\definecolor{light-pink}{rgb}{1,0.5,.5}

\usepackage{listings}
\definecolor{codegreen}{rgb}{0,0.6,0}
\definecolor{codegray}{rgb}{0.5,0.5,0.5}
\definecolor{codepurple}{rgb}{0.58,0,0.82}
\definecolor{backcolour}{rgb}{0.95,0.95,0.92}

\lstdefinestyle{mystyle}{
  basicstyle=\ttfamily\footnotesize,
  breakatwhitespace=false, 
  breaklines=true,                 
  captionpos=b,                    
  xleftmargin=.01\textwidth, xrightmargin=.01\textwidth,
  numbers=left,                    
  numbersep=5pt,                  
  showspaces=false,                
  showstringspaces=false,
  showtabs=false,                  
  tabsize=1,
  frame=single,
}
\makeatletter
    \newcommand{\linebreakand}{%
      \end{@IEEEauthorhalign}
      \hfill\mbox{}\par
      \mbox{}\hfill\begin{@IEEEauthorhalign}
    }
\makeatother

\makeatletter
\patchcmd{\@maketitle}
  {\addvspace{0.5\baselineskip}\egroup}
  {\addvspace{-2\baselineskip}\egroup}
  {}
  {}
\makeatother

\begin{document}

% \title{An Empirical Analysis on Virtual Reality Application Testing}
\title{ Characterizing Virtual Reality Software Testing}

% \author{
% \IEEEauthorblockN{Dhia Elhaq Rzig\IEEEauthorrefmark{1},
% Nafees Iqbal\IEEEauthorrefmark{2}}
% \IEEEauthorblockA{University of Michigan - Dearborn\\
% Dearborn, MI, USA\\
% \IEEEauthorrefmark{1}dhiarzig@umich.edu,
% \IEEEauthorrefmark{2}nafees@umich.edu}
% \and
% \IEEEauthorblockN{Isabella Attisano\IEEEauthorrefmark{3},
% Xue Qin\IEEEauthorrefmark{4}}
% \IEEEauthorblockA{Villanova University\\
% Villanova, PA, USA\\
% \IEEEauthorrefmark{3}iattisan@villanova.edu,
% \IEEEauthorrefmark{4}xue.qin@villanova.edu}
% \linebreakand
% \IEEEauthorblockN{Foyzul Hassan}
% \IEEEauthorblockA{University of Michigan - Dearborn\\
% Dearborn, MI, USA\\
% foyzul@umich.edu}
% }

\author{ \IEEEauthorblockN{Dhia Elhaq Rzig}
  \IEEEauthorblockA{University of Michigan - Dearborn\\
  Dearborn, MI, USA\\
dhiarzig@umich.edu}
\and
 \IEEEauthorblockN{Nafees Iqbal}
  \IEEEauthorblockA{University of Michigan - Dearborn\\
    Dearborn, MI, USA\\
nafees@umich.edu}
\and
\IEEEauthorblockN{Isabella Attisano} 
  \IEEEauthorblockA{Villanova University\\
   Villanova, PA, USA\\
iattisan@villanova.edu}
 \linebreakand
\IEEEauthorblockN{Xue Qin}
\IEEEauthorblockA{Villanova University\\
 Villanova, PA, USA\\
xue.qin@villanova.edu}
\and
\IEEEauthorblockN{Foyzul Hassan}
\IEEEauthorblockA{University of Michigan - Dearborn\\
Dearborn, MI, USA\\
foyzul@umich.edu}}


% \affiliation{%


\maketitle
\bibliographystyle{IEEEtran}
\begin{abstract}
%%\todo{Assigned to: Foyzul}\\
Virtual Reality (VR) is an emerging technique that provides a unique real-time experience for users. VR technologies have provided revolutionary user experiences in various scenarios (e.g., training, education, product/architecture design, gaming, remote conference/tour, etc.). However, testing VR applications is challenging due to their nature which necessitates physical interactivity, and their reliance on hardware systems. Despite the recent advancements in  VR technology and its usage scenarios, we still know little about VR application testing. To fill up this knowledge gap, we performed an empirical study on 97 open-source VR applications including 28 industrial projects.
% to answer the following questions: (1) to what Extent are test cases developed for VR applications, (2) how effective are the test cases developed in VR applications, (3) what is the design quality of test cases developed for VR applications (4) what is the Taxonomy of VR projects’ Test cases. 
Our analysis identified that 74.2\% of the VR projects evaluated did not have any tests, and for the VR projects that did, the median functional-method to test-method ratio was low in comparison to other project categories. Moreover, we uncovered tool support issues concerning the measurement of VR code coverage, and the code coverage and assertion density results we were able to generate were also relatively low, as they respectively had averages of 15.63\% and 17.69\%. 
% In addition to a lack of prevalence and effectiveness, VR test cases suffer from test quality issues such as the Assert Roulette test smell, which was present on average within 38.43\% of a project's tests. Finally,
% % Eager Test, affecting an average of 20.52\% of a project's tests, among others.
% through manual testing-characteristics analysis on 220 VR test cases, we developed a VR test objective taxonomy with eleven main categories, and a comparitve analysis on 291 non-VR test cases allowed us to determine which categories were specific to VR projects.
% We believe our results constitute a call to action for the VR development community, alerting them to various issues with their current testing approaches and providing them guidelines to the various testing categories, especially VR-specific testing categories, they need to target. Furthermore, our findings present the opportunity for further research on pattern-based automated test case generation and test transplantation for VR projects, and provide general guidelines through our test taxonomy.  testing-characteristics
Finally, through manual analysis of 220 test cases from four VR applications and 281 test cases from  four non-VR applications, we identified that VR applications require specific categories of test cases to ensure VR application quality attributes. We believe that our findings constitute a call to action for the VR development community to improve testing aspects and provide directions for software engineering researchers to develop advanced techniques for automatic test case generation and test quality analysis for VR applications.

% We also observed that VR test cases show commonality for VR-specific test objectives, which can be an opportunity for further research on pattern-based automated test case generation and test transplantation for VR projects. 
% Our replication package, containing the dataset we used, software tools we developed, and the results we found, is accessible at~\url{https://figshare.com/s/7e3aa28b1415d26a7222}.
% \listoftodos \\

\end{abstract}



% \keywords{Empirical Study, Virtual Reality, Software Testing}

\newcommand{\totalCountTestClasses}{523}
\newcommand{\totalCountTestMethods}{3051}
\newcommand{\totalCountTestLOCs}{68939}
% \vspace{-0.15cm}
\section{Introduction}
\vspace{-0.27cm}
%%\todo{Assigned to: Xue}
% 1. The growing number of VR apps development makes the quality control a need.
% TODO: Dhia: add we did not separate Game and Non-Game for both VR and Non-VR.
Virtual Reality (VR) applications provide an immersive experience to end-users with a computer-generated environment that includes scenes and objects that appear real in their surroundings. 
Although the term virtual reality was introduced three decades ago~\cite{Rheingold}, its real surge started around 2016, with the release of VR devices such as Oculus Rift~\cite{oculusRift} and HTC Vive~\cite{Vive}, and software support from the likes of Unity and Unreal Engine. Indeed, both industrial and personal usage have surged in recent years. According to a 2021 study~\cite{vr_market_2022}, the global virtual reality market exhibited a significant growth of 42.3\% in 2020 compared to the years 2017-2019, and this market is projected to reach over \$80 billion within the next seven years.
To accurately simulate the user experience and to support areas such as education, product/architecture design, remote conferencing/touring, and surgical procedures, high-quality VR software is essential. However, existing automated techniques' support for VR software development is still at an early stage with few available tools and frameworks, especially for VR testing.

% Virtual Reality (VR) Application is an emerging field that provides a unique real-time experience for users. 
% With applications in areas such as education, product/architecture design, gaming, remote conference/tour, and surgical practicing, 

% By 2027, the VR market is expected to reach \$101.9 billion, with a 37.4\% growth in popularity and demand. VR technologies have provided revolutionary user experience. To deliver such services, it is crucial to develop high-quality VR software. 

% 2. So far, what are the state-of-art tools in software testing? in VR testing?
% Any empirical study regards the testing in mobile? normal software? VR?
% What are the existing study in VR?
% conclude that very little study has been done to study the testing under VR context. 

The importance of software testing and quality assurance has been widely affirmed in both academic and industrial communities. 
To address the challenges and design the appropriate solutions, researchers have carried out many testing-practices studies with different approaches including human-developers-oriented interviews~\cite{Greiler2012} and large-scaled quantitative and qualitative empirical studies~\cite{Kochhar2013, Memon2017}. 
Moreover, test-practice investigations have been reported for different application types: mobile applications~\cite{Pecorelli2020TestingOM, Peruma2019}, Machine learning applications~\cite{Wang2021}, and many more~\cite{Bowes2017,Palomba2017OnTD,Palomba2017DoesRO,Nejadgholi2019,cadar2011symbolic}. 
Testing VR applications is a challenging task~\cite{Ashtari2020,zhang2020virtual,correa2018automated} due to the complex design of VR projects, the user-immersive experience, and inadequate technical support for VR development, debugging, and testing activities.
So far, existing VR research activities have focused on development support, such as performance optimization~\cite{Nusrat_2021}, 
code dependency~\cite{Jacinto_2021}, and code smell detection~\cite{BorrelliACM2020}. 
In addition, there are several studies on Game applications, such as regression testing for Games~\cite{Wu2020ICMSE}, and a differentiation study between Game and Non-Game applications~\cite{Pascarella2018}.
However, none of the existing works studied the prevalence, effectiveness, design quality, and characteristics of VR software testing.
% However, very little has been studied about the effectiveness, design quality, and characteristics of VR software testing. VR application testing is challenging due to its different characteristics, such as real-time user interactions and stimmulated environments. 
% While traditional software usually has many powerful testing tools~\cite{} and well-designed strategies~\cite{} to support and speed up the testing, most VR projects are tested manually late in the development stage. This method is labor intensive and might lead to failures in software delivery due to late-stage testing. 
% 3. Discuss why this type of study is important, and what are the potential benefit
To remedy this knowledge gap, we opted to perform a qualitative and quantitative analysis of the existing testing practices in VR applications. 
% This type of analysis was a challenging but rewarding process.
% This analysis empowered us to provide empirically-grounded results to researchers and developers, which will help them identify the procedures they need to support or improve  to assure a successful application release. 
% 4.To fill the gap, what are the research questions the paper try to answer?
% Considering the aforementioned reasons, 
We carried out an empirical study on 97 VR applications in this paper, ranging from small-scale projects to projects backed by large companies and organizations like Microsoft and Unity Technologies, where we analyzed the prevalence, quality, and effectiveness of existing VR tests. Then, we categorized the different VR tests into different types based on their characteristics.Finally, we analyzed the tests of four hand-picked non-VR projects to determine which of the aforementioned test categories were VR-specific. 
% In the following, we will specify four main research questions we want to address in our study and briefly discuss the motivation and findings for each.
% \begin{itemize}
% \item \textbf{$RQ_1$:} To What Extent Test Cases Are Developed for VR Applications?
% \item \textbf{$RQ_2$:} Effective of Test Cases Developed in VR Applications
%     \item \textbf{$RQ_3$:} Design Quality of Test Cases Developed for VR Applications
%     \item \textbf{$RQ_4$:} What are the types of Test Cases Developed in VR Applications 
% \end{itemize}
% \vspace{-0.1cm}

\noindent Our main research questions are: 
\begin{itemize}[leftmargin=.1in,nolistsep]
    \item \textbf{RQ1}: To What Extent Are Test Cases Developed for VR Applications? \\
    \textbf{Motivation.} This question allowed us to estimate the current effort that VR developers put into testing their projects and understand the potential need for VR testing support. \\
    \textbf{Answer.} We discovered test cases in only 25 out of the 97 VR projects we analyzed. Moreover, we found that 
    % the medians of the test-to-code  method ratio and class ratio were
    % 0.048 and 0.054. 
    % A recent study~\cite{Vidacs2018} on co-evolution analysis of production and test code suggests functional classes and test methods are added in parallel with a ratio of around 1.33.
    the current ratio of code-to-test is too low using both class and method granularities, especially as they were more than 26 times lower than those found by Vidacs et al.~\cite{Vidacs2018} who performed a similar analysis on test-to-code ratios. We believe these results indicate an important need for improved testing support. 

    \item \textbf{RQ2}: How effective are the test cases developed in VR applications? \\
    \textbf{Motivation.} To develop a deeper understanding of the status quo of VR testing, we adopted two state-of-the-art metrics to evaluate the effectiveness of the existing test methods: code coverage and assertion density. \\
    % These reports can provide future insights for test design improvement. \\
    \textbf{Answer.} For code coverage, within the 7 measurable projects,
    % the highest rate is 31.5\%, the lowest rate is 1.29\%, and the average rate is 15.63\%. 
    the average rate was more than 5 times lower than the recommended rate of 80\%~\cite{Atlassian,Hemmati2015}, and the assertion density has values that are lower than the recommended rates, and which are linked to higher bug rates within software projects~\cite{Kudrjavets_Nagappan_etal_2006}.
    These findings indicate that the testing practices of VR projects are less effective. 

    % For assertion density, the values vary from 22.17\% to 0.00\% with a median value of 5.96\%. This is in contrast to prior studies~\cite{Atlassian,Hemmati2015} which recommend more than 80\% of code coverage for quality attributes maintenance.
    
    \item \textbf{RQ3}: What is the design Quality of test cases developed for VR applications? \\
    \textbf{Motivation.} The metrics we used in RQ2 do not reflect the design quality of test methods.
    With this question, we wanted to evaluate the quality
    % discover the test smells 
    of existing tests, and we believe these findings can help guide future testing tool design.\\
    \textbf{Answer.} Using the work of De Bleser et al.~\cite{de_bleser_2019} as a guide, we analyzed the tests for 6 smell types. We found that on average 38.43\% of a project's tests have at least 1 smell type, and a project can have as much as 92\% smelly tests.
    % and report the scores of
    % Within 25 projects with tests, 
    % 18\% of them have Assertion Roulette within 20\% or more of their test cases, and the worst rates can be as high as 91.96\%.
    % General Fixture was found within 18\% or more of the test cases of 12 projects. 
    % Eager Test smell was found in 10\% or more the tests of the 17 projects. 
    % Mystery Guest is less common with its presence in 14\% or more of the tests of only three projects;
    % Lazy Test and Sensitive Equality were not found in any of the projects. 
    % To further validate the automated report, two co-authors have labelled 220 test methods.
    % After the comparison with manual results, automatically results show 91.35\% accuracy score with 92.62\% recall and 91.98\% F-1 score.
    This indicates that test smells are common within most VR test methods, resulting in a lower design quality.

    \item \textbf{RQ4}: What are the different categories of VR Test Cases and which categories are specific to VR? \\
    \textbf{Motivation.} With this question, we aimed to discover testing scenarios that reflect
    % the nature and 
    characteristics of VR applications. VR applications differ from other application types because of their specific-hardware support, new user experiences, and unique immersive design, among other characteristics. Similar to Muccini et al.'s work~\cite{Muccini_2012} which examined the differences between mobile and traditional applications, we aim to fill a similar knowledge gap and  provide insight for future VR testing design.
      \\ 
    \textbf{Answer.} In our study, we manually analyzed 220 test methods from 4 VR projects that had at least a 10\% code coverage. In total, we  defined 10 main testing categories, such as \textit{Physics Test}, \textit{Animation Test}, \textit{Graphics Test}, \textit{Asset Test}, etc., which are detailed within~\autoref{sub:sec:taxonomy}. To further understand which categories are specific to VR projects, we followed the same methodology and analyzed 281 test methods from four non-VR projects that meet the same code coverage criteria.
    We found four main categories and four subcategories of two main categories were not present in any of these non-VR test methods.
%    And we also present an code illustration for all VR related test types. 

% Our findings to fill the knowledge gap regarding the current state of VR projects' testing, the different categories of VR-testing and the differences between the VR and non-VR testing. Our findings further provide insight to the research community and inspire new and updated designs for VR test automation via the test patterns we uncovered.
% We believe the test patterns we discovered can help researchers' future exploration of test automation.
\end{itemize}

% % 5. What are the methodology and major findings 
% Our automatic analysis shows that \todo{summary of the RQ1 to RQ3}

% To have a better understand of the test design in VR apps, we manually analyze 314 test cases from top four apps with at least 10\% test code coverage.
% In this study, we focus on categorize tests that reflect the nature of the Virtual Reality. 
% More specifically, the Unity APIs which support the features like collision handling, motion validation, camera detection, etc. 
% We further define \todo{XXX} different categories to systematically represent these various of characteristics, including: \todo{list a few examples of categories} 

% 6. contributions
% \vspace{1mm}
\noindent The contributions of this paper are:
\begin{itemize}[leftmargin=.1in,nolistsep]
    \item The first quantitative and qualitative study on existing test practices of VR applications
    % , where we investigated 97 Unity VR projects with C\# programming language. % to Comprehensive investigation on the existing test practice on 97 VR applications.
    \item The first tool for test effectiveness analysis and test smell identification for Unity-based projects. 
    \item A detailed test case effectiveness analysis via the test coverage and assertion density metrics, and detailed test-quality analysis through the test-smell detection.
    % The results show a vast need for test effectiveness and quality improvement. 
    \item A taxonomy containing 10 main test categories which reflect the characteristics of the VR applications, 
    % resultant from our manual analysis on 220 test cases,
    as well as the identification of VR-specific categories within this taxonomy. 
    % It can inspire future research on VR-test automation techniques.
\end{itemize} 

% 7. structure of the paper
This paper is organized as follows: Section~\ref{sec:background} presents the background that defines terms we later use in our manual analysis;
Section~\ref{sec:approach} contains details about the dataset and the methodology used within our automated and manual analysis;
Section~\ref{sec:evaluation} describes the evaluation which answers all four research questions, followed by threats to validity in Section~\ref{sec:discussion};
Section~\ref{sec:related} includes related work; Section~\ref{sec:implications} describes the implications of this paper and we conclude the work in Section~\ref{sec:conclusion}.
% \vspace{-0.4cm}
\section{Background}
\label{sec:background}
\vspace{-0.15cm}
%%\todo{Assigned to: Foyzul}
Automatic software testing allows developers to test application code in an automated, rapid, and reliable way. Similar to traditional software applications, automated software testing can also be applied to VR applications. 
In this study, we analyzed the test characteristics of 97 Unity-based VR applications collected from UnityList~\cite{unitylistbib}, since Unity is one of the most popular frameworks for developing VR applications~\cite{Doucet_2021}. 
% In this section we will discuss the automatic testing approach of VR applications and their related terminologies that are interconnected with VR testing process.
% However, automated testing of VR applications is currently limited to unit testing. 
VR tests mainly focus on the behavior of VR subsystems and class components.
While analyzing VR tests, familiarity with how Unity works and some terms is required. 
% These details are presented within the following paragraphs.

% \noindent\textbf{Scenes:} a scene is a type of asset where a developer works with content within Unity. A game may have one or multiple scenes, within which  environments, characters, obstacles, and other assets may be loaded and used. The objects most commonly present and manipulated within a scene are GameObjects. 

\noindent\textbf{Physics System:} 
ensures that the virtual objects correctly respond to different forces such as collisions
and gravity.
Unity platform provides Rigidbody APIs to enable the physics engine control of objects.
Collision and Colliding are also important concepts in VR projects that define how virtual objects react to overlapping with or without physical effects.
% For example, a character may pass through a fog object, and collide with a wall object.
% In general, Unity handles the events by following the time order: at the beginning of the event, during the event, and on leaving the event. 

\noindent\textbf{Graphics System:} 
enables developers to control the appearance of VR applications. This includes Rendering, Display, Camera, Lighting, etc.
In 3-D graphic design, rendering is the process of adding shading, color, and lamination to a 2-D or 3-D wireframe in order to create life-like images on a screen. 
% Rendering is to make a life-like virtual environment by adding shading, color, etc..
This process can be preloaded or occur in real time when users interact with VR applications. 
Display is related to displaying the rendered objects within the VR scene, which users can view through hardware such as monitors or head-mounted devices.
Unlike non-VR applications, VR applications can create multiple cameras in the same scene, and the display will update along with the camera switching and location changing. The camera represents the view angle that the user utilizes to see the virtual world.

\noindent\textbf{Animation System:}
allows developers to animate target objects via jumping, moving, stopping, rotating, etc.
The animation design in VR applications is more complicated than that of non-VR applications.
With fixed angles in non-VR applications, animation design is a linear process that focuses on the representation from a locked direction or view.
However, in VR applications, animating user surroundings is a parallelized process. Developers need to ensure the correctness of the representation from any arbitrary angle.

\noindent\textbf{Other terms:}
\textit{GameObject:} the fundamental class for all the objects in a virtual world. By combining the different controls and features, developers can realize customized functions like moving objects.
\textit{Colliders} represent the invisible physical shapes of objects. The Physics system uses them to decide the physical effect like overlapping between objects.
\vspace{-0.2cm}
\section{Research Approach}
\vspace{-0.15cm}
\label{sec:approach}

% \vspace{-0.4cm}
\begin{figure}[!htbp]
    \centering
     \includegraphics[width=0.92\linewidth]{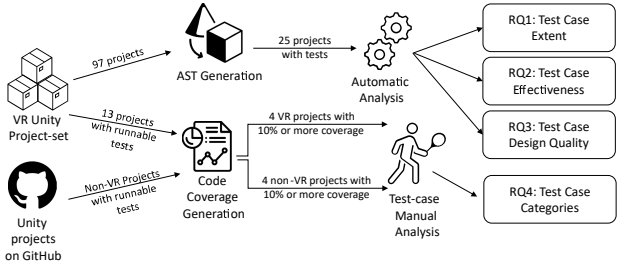}
    \caption{Overview of Research Approach}
    \label{fig:overall}
    \vspace{-0.4cm}
\end{figure}
% \vspace{-0.3cm}
The research approach overview is illustrated within~\autoref{fig:overall}. In this section, we will first introduce the studied dataset, then describe the AST-based automatic analyses used to measure the test prevalence, efficiency, and quality, and eventually discuss the manual analysis used to discover the taxonomy of VR tests as well as identify VR-only test categories. 

\subsection{Dataset}
\vspace{-0.15cm}
\label{sub:sec:dataset}
Within this work, we wanted to study a group of Open Source Unity-based Virtual Reality software projects. Since collecting a dataset can be a time-consuming process and is not a goal of this study, we opted to use the dataset of Nusrat et al.~\cite{Nusrat_2021}, containing 100 VR projects which are Unity-based. We specifically chose this dataset as it contains manually-verified VR projects from Unity, one of the most popular Game and VR development engines~\cite{Doucet_2021}. We were unable to obtain 3 projects due to de-listing. We identified the different versions of Unity used in the development of the projects we obtained: 3 projects were using Unity 2021, 1 project was using Unity 2020, 18 projects were using Unity 2019, 24 projects were using Unity 2018, 23 projects were using Unity 2017, 27 projects were using Unity 5, 1 projects was using Unity 4. This diversity of VR projects should allow us to uncover knowledge about testing practices and tool support across various generations of the Unity framework. Furthermore, while the majority of the projects we considered were independent and academic projects, respectively 55 and 14 projects, it also contained 28 industrial projects backed by the companies and organizations like Microsoft, Unity Technologies, and Vive. All VR projects contain a minimum of 100 commits, with average and median commits of 728.42 and 204, respectively. 
% They also have a wide range of ages, ranging from 2 to 2363 days, with an average of 567.74 days and a median of 333.5 days. 
Furthermore, they are composed of teams ranging from 1 person to 1568 people, with an average of 61.28 and a median of 4 people.   
% Finally, the projects cover a wide range of ages, team size, and activity as measured by number of commits. These properties are detailed within~\ref{table:proj-props}.

In order to be able to compare and contrast the types of tests used within non-VR projects and those used within our set of VR projects, we manually collected four Unity-based non-VR projects, based on similar criteria to those used by Nusrat et al., and verified that they had a code coverage of at least 10\%, similar to that of the four VR projects we selected for the manual analysis and test taxonomy construction, which are detailed within~\autoref{sub:sec:method_manual_analysis}. The four VR projects we selected for manual analysis were composed of two Game and two Tool projects. To take into account of  characteristics these two project types during our comparative analysis, we selected two non-VR Game projects, and two non-VR Tool projects for our non-VR comparison set.  

% \begin{table}[H]
% \small
% \centering
% \begin{tabular}{|c||c|c|c|c|c|c|c|} 
%  \hline
%  & Age & Team  & Nb.  \\
%  & In Days & Size & commits 
%  \\ 
%  \hline\hline
%  Min. &  2&	1&	100 \\ 
%  \hline
%  Max.& 2363	&1568&	17066 \\
%  \hline
%  Avg. & 567.74 &	61.28&	728.42 \\
%  \hline
%   Median & 333.5&	4&	204\\
%  \hline
% \end{tabular}
% % \vspace{1mm}
% \caption{Properties of the VR-set projects}
% \label{table:proj-props}
% \end{table}
% TODO: add number of indie vs industry projects + info about projects 
% \vspace{-0.2cm}
\subsection{Methodology of Automatic Analysis}
\vspace{-0.15cm}
\subsubsection{Static Analysis of Unity Projects}
% Describe how unity projects are developed in C\#, decision to use srcml, etc. 
Since Unity  makes use of the standard .Net Framework, alongside its internal frameworks and classes, and the C\# programming language for its scripting~\cite{UnityArch}, we needed to use an AST generator that supported these technologies.
% While it is possible to use Roslyn~\cite{roslyn}, the C\# compiler in order to generate and parse C\# code ASTs, this tool can not compile Unity C\# code~\cite{unity_vs}, and thus may cause issues in the generation of corresponding ASTs. 
We opted to use SrcML~\cite{srcML} to generate the ASTs of the Unity C\# code. SrcML is a research tool that allows the generation of ASTs for various programming languages, which facilitates the extension of our approach and tools to other sets of projects that use different programming languages. It supports C\#, and does not rely on compilation to generate ASTs, thus allowing us to avoid any compatibility problems. For each VR project within our dataset, we analyzed its repository to extract the C\# code files and then generated the AST of each file to perform the different analyses described in the following sections. 

% \vspace{-0.3cm}
\subsubsection{VR Test Cases Prevalence}

% Describe number of test cases found per project, ratio of test cases to functional code 
In order to evaluate the prevalence of test cases within our set of VR projects, 
% first, we programmatically generate the ASTs for all the methods based on the approach discussed above.
% Then, 
We counted the number of the test methods and
% which are labeled \texttt{[Test]} or \texttt{[UnityTest]},
the test classes, in addition to the number of functional methods and classes of our VR projects. 
% In order to detect test cases within our dataset of Unity projects, we used the AST generation method described above, and then programmatically analyzed the different ASTs to find and count the methods labeled \texttt{[Test]} or \texttt{[UnityTest]}, the two labels used to delineate test methods within Unity projects, as well as the count of the classes containing these methods which are considered test classes, and the LOCs of these methods and classes. In addition, we also extracted the counts of methods, classes and LOCs of the functional code of our VR projects.
% It's important to note that we configured our code to ignore libraries or packages imported by the different Unity projects, as they do not form part of the code programmed and maintained by the projects' maintainer, and but are rather distinct functional and test code imported by the Unity projects we're analyzing.
Using these data points, we calculated the following metrics for each project: \\

% \noindent\textbf{Ratio of Test LOC Count to Functional LOC Count:}
% \vspace{-1mm}\\
%  \resizebox{0.7\hsize}{!} {
% $TestToCodeLOCRatio=  \frac{\sum{}^{}LOC(TestMethods)}{\sum{}^{}LOC(FuncMethods)}$
% }
% \vspace{1mm}

% \noindent\textbf{Ratio of Test Method count to Functional Method count:}
\begin{center}
\vspace{-5mm}
\resizebox{0.77\hsize}{!} {
$TestToCodeMethodRatio= \frac{Count(TestMethods)}{Count(FuncMethods)}$
}
\vspace{1mm}

% \noindent\textbf{Ratio of Test Class to Functional Class:}
\vspace{1mm}
\resizebox{0.78\hsize}{!} {
$TestToCodeClassRatio=  \frac{Count(TestClasses)}{Count(FuncClasses)}$
}
% \vspace{0.5mm}
\end{center}

Based on the work of Klammer et al.~\cite{Klammer_Buchgeher_etal_2018} and Williams et al.~\cite{Williams_Kudrjavets_etal_2009},
these  metrics adequately represent the relative frequency of test code within our VR projects.
% These different metrics allowed us to estimate the relative frequency of test code within our unity projects, and was used within comparable works such  as those of Klammer et al.~\cite{Klammer_Buchgeher_etal_2018} and of Williams et al.~\cite{Williams_Kudrjavets_etal_2009}.
\subsubsection{VR Test Cases Effectiveness}
\label{sub:sub:sec:test_effectiveness}
%use android paper for reference for code cov and assert density
A practical way of measuring the effectiveness of tests in a software project is to calculate the Code Coverage~\cite{Gopinath_Jensen_etal_2014}, which denotes the degree to which the functional code is executed after a test suite finishes running. It is calculated via the ratio  of the LOCs of functional code executed when tests are run to the total amount of coverable code within a project.
This metric has been used by Pecorelli et al. in the context of Open-source Android apps~\cite{Pecorelli2020TestingOM}, Ivanković et al. in the context of software projects at Google~\cite{Ivankovic_Petrovic_etal_2019}, and Kochhar et al. also applied it to OSS projects ~\cite{Kochhar_Thung_etal_2015}.
% We believe this widespread adoption proved its validity and credibility in evaluating our dataset.
The equation of code coverage is shown below: \\

%\noindent\textbf{Code Coverage Equation:}
\vspace{-5mm}
\begin{center}
\resizebox{0.9\hsize}{!} {
$CoveragePercentage=\frac{FuncLOCsExecutedWhenTestsRun}{CoverableFuncLOCs}$ 
}    
\end{center}
% \vspace{0.5mm}
% \\

The generation of this metric and coverage reports is only possible through the use of Unity 2019.3 or newer~\cite{unitycodecov}, making it obligatory to upgrade the projects using an older version of Unity to at least that version in order to generate code coverage metrics. 
% Although Visual Studio allows the execution and debugging of Unity projects by attaching to the Unity editor~\cite{unity_vs}, it does not support running tests or generating code coverage reports related to Unity projects. 
We planned to directly generate code coverage reports for 16 projects without upgrading and indirectly generate the code coverage reports for the rest of the projects with tests after upgrading them to Unity 2019.3 or newer. 
However, out of the 25 projects that we detected had tests via our automatic method, only 13 had tests runnable via Unity. In addition, we were only able to generate code coverage reports for 7 out of these 13 projects. 
Among the projects for which we were unable to generate coverage reports, \texttt{addyi@SoftwareCity} and \texttt{RussellXie7@Unity\_Hololens\_Dev}, were using Unity 5, and their tests were no longer runnable via Unity when they were updated to Unity 2017 or newer. Three projects had various compilation problems. Finally, \texttt{Microsoft@MixedRealityToolkit-Unity}, runs the coverage tool but its report is not generated. 
% The generation of these metrics and coverage report in general is only possible through the use of Unity 2019.3 or newer~\cite{unitycodecov}, making it obligatory to upgrade the projects using an older version of Unity to at least that version in order to generate code coverage metrics. It's important that Visual Studio only allows the execution and Debugging of Unity projects, by attaching to the Unity editor~\cite{unity_vs}, and does not support running tests or generating code coverage reports related to Unity projects. Theoretically, we could directly generate code coverage reports for 16 of our project-set, and indirectly generate the code coverage reports for the rest of the projects after upgrading them to Unity 2019.3 or newer. Out of the 62 projects we detected had tests via our automatic method, 13 had tests which were runnable via Unity. We were able to generate code coverage reports for only 7 out of them. Among the projects not allowing us to generate test coverage report, two projects: \texttt{addyi/SoftwareCity} and \texttt{RussellXie7/Unity\_Hololens\_Dev} were developed with Unity 5 and their tests were no longer runnable via Unity when they were update to Unity 2017 or newer, 3 projects had various compilation problems, and 1 project, \texttt{/microsoft/MixedRealityToolkit-Unity} would run the coverage report tool but no report is generated or saved. 

To further clarify our findings regarding the effectiveness of the tests from our VR project set and circumvent some of the issues we encountered for  code coverage report generation, we opted to use the Assertion Density~\cite{Kudrjavets_Nagappan_etal_2006,Catolino2019} metric to evaluate the test cases. This metric is calculated via the ratio of the number of assertions to the length of test cases that contain them. This metric's equation is: \\

%\noindent\textbf{Assertion Density Equation:}
\vspace{-5mm}
\begin{center}
\resizebox{0.63\hsize}{!} {
$AssertionDensity=\frac{NbAssertions}{TestLOCs}$
}
\end{center}
% \vspace{-mm}

\subsubsection{VR Test cases Design Quality}
\label{sub:sub:sec:test_smells}
% describe the 6 test smells, subsection for reach smell + how it's programmatically detected 
In order to evaluate the quality of the test cases within our project set, we scanned them for Test smells~\cite{Meszaros_2007,Deursen01refactoringtest,Tufano_Palomba_etal_2016}. They are similar to regular code smells in being symptomatic of technical debt and predicting future problems, but they are specific to test code. We considered six test smells from the work of De Bleser et al.~\cite{de_bleser_2019}, as they were found to be the most prevalent in a collection of previous works~\cite{Greiler_vanDeursen_2013,
Greiler_vanDeursen_2013,
Palomba_2018}, and adapted them to the context of VR applications. These smells are: 

\noindent\textbf{Assertion Roulette (AR)}: If a test case contains more than one assertion, at least one of which does not provide a message when detecting a failure, it can be hard to diagnose which problems are present within the functional code being tested. 
% To detect this smell, our tool first generates the ASTs of test code classes, then scans each method within these ASTs to find if they contain more than one assertion without messages. If so, the tool will calculates their ratio in comparison to the total number of assertion statements in that method.

\noindent\textbf{General Fixture (GF)}: A test fixture is too general if it initializes fields that are not used by one or more test methods, making it difficult to discern which fields are being shared by the different test methods within a test class.
% Our tool analyzes the AST of each test class to extract the fields that are initialized by the \texttt{@Setup} or equivalent text fixture, as well as the fields that are used by each test methods. The tool then calculates the ratio of tests that leave one or more fields unused in comparison to the total number of tests.

\noindent\textbf{Sensitive Equality (SE)}: If a test case has an assertion that compares the state of objects by comparing their representations as text, for example by using their ToString() methods, it makes itself susceptible to errors due to irrelevant textual representation details such as spaces.
% Our tool detects this smell by analyzing the comparisons within each test method to see if they are comparing two objects using their textual representations directly or indirectly via intermediate variables. Finally, the ratio of the test methods containing these smells in comparison to all the test methods of a project is calculated.

\noindent\textbf{Eager Test (ET)}: We consider a test that evaluates more than one functional code method with the same fixture Eager. This smell violates the principle that every test case should only test one method, and the test failure should only signify issues within that method, not another irrelevant method. 
% In order to detect this smell, our tool generates the ASTs of Test code classes, then scans the assertion statements of each method within them. Later, it extracts the different method calls which represent the parameters of these assertions, and if it finds that more than one functional code method is being called within them, the test method will be marked as eager. Finally, the tool calculates the ratio of Eager tests out of all the tests of a specific project. 

\noindent\textbf{Lazy Test(LT)}: 
A test case is lazy if it tests the same functional method using the same fixture as another test method.
% as another test case or if its test results depend on other test cases.
The problem this smell implies is that
% that any updates that have been applied to one test method will affect other test methods. Moreover, 
after modifying one function that is being tested, multiple test methods need to be updated accordingly. Thus this smell will affect the maintainability of test cases.
% A test case is considered lazy if it's testing the same method as another test case, and if it's results can be modified depending on the execution of other test cases, thus affecting the maintainability and predictability of the test cases. The problem this smell implies is that you have to modify multiple tests, instead of a single one when modifying that same method, and that changing the order of the tests can change their respective results. 
% To detect this smell, our tool generates the ASTs of Test code classes, then scans the assertion statements of each method within them. Later, it extracts the different method calls which represent parameters of these assertions, and 
% To detect this smell, our tool first scans the assertion statements of each method within the generated ASTs of test code classes, then extracts the different method calls which represent the parameters of these assertions. 
% If it finds that the same functional code method is being called within assertion statements of different tests, or if an object generated by a method during one test is being used by a separate test, the test methods are marked as Lazy. Finally, the tool calculates the ratio of Lazy tests out of the all the tests of a specific project. 

\noindent\textbf{Mystery Guest(MG)}: A test case has this smell if it uses external resources that are not managed by a fixture or are not Mock objects. This smell may cause issues since external resources might change over time or be unavailable during test-case execution. For example, a test method can fail if a specific database (\textbf{sqlite} object) or \textbf{file} object are not available during it's execution.
% This smell makes the test case nondeterministic by possibly causing test failures due to reasons unrelated to the tested functions. 
% To detect this smell, our tool generates the ASTs of test code classes, then scans the objects being used within each test method. Later, it extracts the different types of these objects to find if they represent an external object, for example, a mystery guest for a test method can be \textbf{sqlite} or \textbf{file} object when they are not initialized as mock objects. It also denotes the usage of local resources, such as assets stored in files, as a method's parameters.

% \vspace{-0.2cm}
\subsubsection{Evaluation of smell-detection tool}
\label{sub:sub:sec:eval_auto}
To evaluate the accuracy of our smell-detection tool and verify the correctness of our findings, we first applied the automated approach to four VR projects from the dataset with a coverage rate above 10\%. Then two co-authors manually evaluated the same projects separately by labeling the test methods with any corresponding smell types. Both co-authors found 220 test methods, which is the same number found by the tool. An average Kappa of 0.92 was found between the authors across the different smell types, signaling high agreement, and any differences were then resolved via discussion. 
Upon evaluating the automatically detected smells using the manual observations as a baseline, on average, an accuracy of 91.35\%, a recall of 92.62\%, and an F-1 score of 91.98\% were found across the aforementioned test smell categories. 
%These categories were Assertion Roulette, Eager Test, General Fixture, and Mystery Guest.
As noted in~\autoref{sub:sec:test_case_quality}, no instances of Lazy Test and Sensitive Equality were found via the manual or automatic analyses we performed. To verify the correctness of our tool for these smells, we developed one test stub for the sensitive equality smell and two test stubs for the Lazy test smell and verified that our tool correctly detects these smells.
% To evaluate the accuracy of our smell-detection tool and verify the correctness of our findings, we automatically analyzed the 4 projects from our project-set that had a coverage rate above 10\%. A total of 220 test methods were found within these projects, and 2 co-authors who manually evaluated these projects found the same number of test methods. The 2 co-authors also evaluated these test methods for the different smells we considered within our work, and an average Kappa of 0.92 was found between them across the different smell types, signaling high agreement, and any differences were then resolved via discussion. Upon evaluating the automatically detected smells using the manual observations as a baseline, an average accuracy of 91.35, an average recall of 92.62 and an average F-1 score of 91.98 were found across the different smell categories of which instances were found. These categories were Assertion Roulette, Eager Test, General Fixture and Mystery Guest. As noted in~\autoref{sub:sec:test_case_quality}, no instances of Lazy Test and Sensitive Equality were found via manual or automatic analysis. To verify the correctness of our tool for these smell, we developed 1 test stub for the sensitive equality smell and 2 test stub of the Lazy test smell and verified that our tool correctly detects these smells.
% \vspace{-0.25cm}
\subsection{Methodology of Manual Analysis}
\vspace{-0.15cm}
% Describe the detail related to the manual analysis and classification of test cases into different types
\label{sub:sec:method_manual_analysis}
%\todo{Nafees}
To explore the characteristics of VR test cases and understand the differences between VR software testing and non-VR software testing, 
we carried out a manual analysis with a focus on discovering unique testing scenarios, exploring the test design patterns, and testing goals.
Based on the automatic results within~\autoref{sub:sec:test_case_effectiv}, we ranked all the VR projects with executable test methods by their code coverage and selected the ones with at least 10\% coverage. We believe this selection allows us to select projects which have relatively more effective test methods than others.
Eventually, we chose four projects, composed of two Tool and two Game projects, and which included 220 test methods.

There were no existing categories for authors to use as a  reference when this study was conducted.
Since we had to start from scratch in our categorization of the VR test types, we designed  our manual approaches
% have been designed in a way
to minimize bias.
First, two co-authors separately observed all the test methods and their related source code, and also performed an exploration of Unity documentation and VR developers' forums to generate a comprehensive report for every test method.
This report included details such as Unity API calls, observed tested target and environment behavior, test scenario description, text method code pattern, and corresponding tested functional code pattern.
Then, we asked a third co-author with previous VR experience to join the process. We assume this author carries less bias in deciding the categorization of the VR test code designs which were uncovered in the earlier step, as this author did not participate in it. These three co-authors then categorized all 220 test methods separately by reviewing the observed records. Eventually, voting within three rounds of consensus meetings was carried out to finalize the results and resolve any disagreements. Before resolving the disagreements, Fleiss' Kappa coefficient was calculated, and was more than 0.81, indicating a very good agreement between co-authors. A similar process of manual categorization and discussion was followed to classify the 281 non-VR test cases, extracted from the four non-VR projects, within the newly-created test taxonomy. The Fleiss' Kappa coefficient for this process was 0.86, indicating very good agreement between co-authors as well. 

\section{Empirical Evaluation}
\label{sec:evaluation}
\vspace{-0.15cm}
\subsection{To What Extent Are Test Cases Developed for VR Applications?}
\label{sub:sec:test_case_extent}
\begin{figure}[!htbp]
    \centering
     \includegraphics[width=0.85\linewidth]{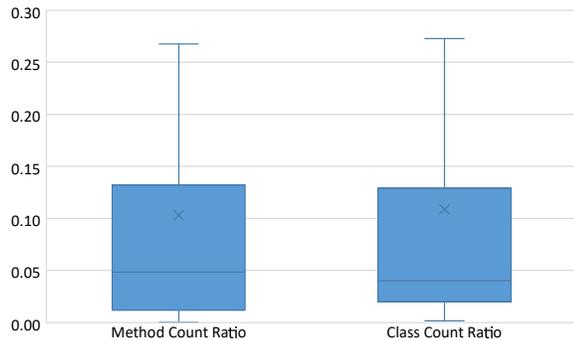}
    \caption{ Method-Count and Class-Count ratios (outliers removed)}
    \vspace{-0.45cm}
    \label{fig:rq1}
\end{figure}

\vspace{-0.15cm}
In order to identify the prevalence of test code in VR applications, we calculated the Method and Class ratios of test code in comparison to functional code for VR projects with test cases.  
% This is especially prevalent through the method count ratio metric, although the class count and LOC ratio metrics don't paint a much rosier picture, since their medians of all 3 are 0.04, 0.048 and 0.054. These medians all drop to 10\textsuperscript{-4} when consider all VR projects, which is unsuprising considering only 25 of the VR projects analyzed had test cases.
% When considering the set of the 25 VR projects with tests, the median values of the LOC ratio, method count ratio, and class count ratio are 0.04, 0.048, and 0.054, respectively. Additionally, these medians all drop to 0 when considering the set of all the VR projects in our dataset. 
% These results are in direct contrast with
The recommended practices indicate that the test code to functional lines of code ratio is 3:1, or in more general cases between 1:1 and 1:10~\cite{loc_ratio}. Indeed, it is recommended that test code be added in parallel to functional code, where one test class should evaluate one functional class, and one test method should evaluate on functional method. This suggests that when considering the Method and Class granularity of ratios, the ideal values are close to 1.
However, the results illustrated in~\autoref{fig:rq1} show that test code only represents a small portion of VR projects' code and that current VR testing prevalence is far from the ideal scenario. 
Indeed, these Method and Class ratios are also significantly worse in comparison to other categories of software projects such as industrial Java projects, analyzed by Klammer et al.~\cite{Klammer_Buchgeher_etal_2018}, where the equivalent LOC ratio was around 0.6 for the totality of the code-base, or C\# projects by Microsoft~\cite{Williams_Kudrjavets_etal_2009}, where the LOC ratio was between 0.35 and 0.89.
% Of the projects which contained test code,
% \texttt{ExtendRealityLtd@Zinnia.Unity} contained the highest values of the ratios calculated. Even though this project is supported by a non-profit, it performed better than  projects developed by for-profit companies, such as Vimeo and IBM.
% Most of the worst-performing projects concerning these metrics are single-contributor projects which have not been updated recently, such as \texttt{myieye/cube-arena} and \texttt{jdknox/vrDemo}. 

% \vspace{-0.3cm}

\subsection{How effective are the test cases developed in VR applications?}
\begin{figure}[!htbp]
    \centering
   \includegraphics[width=0.65\linewidth]{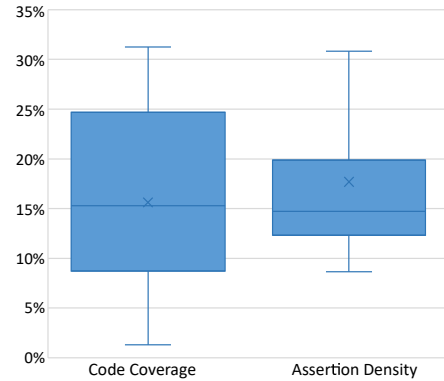}
    \caption{Code Coverage and Assertion Density (outliers removed)}
    \label{fig:rq2}
\end{figure}

\vspace{-0.15cm}
\label{sub:sec:test_case_effectiv}
As discussed within~\autoref{sub:sub:sec:test_effectiveness}, generating code coverage on the majority of Unity projects has proven challenging. Furthermore, the code coverage results we were able to generate, illustrated within~\autoref{fig:rq2}, paint a grim picture of VR testing's effectiveness. While there is no universal code coverage rate that all software projects should target, an 80\% rate of code coverage is the general recommendation~\cite{Atlassian,Hemmati2015}. Yet, the highest coverage rate we noted was 31.25\%, the lowest was 1.29\%, and the median was 15.28\%, thus putting into doubt the effectiveness of the tests within the project for which we were able to measure the code coverage. Furthermore, the issues regarding the lack of code coverage detection for Unity versions below 2019.3, especially when combined with the plethora of problems associated with the process of updating Unity projects using older versions to this version, highlight the lacking tool support in Unity for testing-related activities. When considering the metric of Assertion density, also illustrated within~\autoref{fig:rq2} for projects that had test cases, it is clear that these values are not more encouraging than those of Code Coverage. Indeed, the median value of 14.73\% we found for this metric in VR applications is even lower than that found within mobile applications~\cite{Pecorelli2020TestingOM}, and the values found for Assertion Density and Code Coverage are linked with less effective testing practices~\cite{marick1999misuse}. 
% The project with the lowest measurable code coverage was \texttt{XRTK@XRTK-Core}, which is the official Mixed Reality framework for Unity. This puts into question whether this tool, meant to be used by other mixed reality projects, is reliable enough for widespread adoption. It is especially concerning since other projects that rely on this framework may be directly influenced by its bugs. The project with the lowest non-zero Assertion Density is \texttt{RussellXie7@Unity\_Hololens\_Dev}, an independently-developed  framework.
Overall, the results we found show a massive need for testing support for independent developers of VR projects.

\vspace{-0.4cm}
\subsection{What is the design quality of Test Cases Developed for VR Applications? }
\label{sub:sec:test_case_quality}
\vspace{-0.15cm}
After evaluating the extensiveness and effectiveness of test cases developed within our project set, we analyzed them to evaluate their quality using the method outlined in~\autoref{sub:sub:sec:test_smells}, and we obtained the following results: 
\vspace{-0.1cm}
\begin{figure}[!htbp]
\centering
   \includegraphics[width=0.85\linewidth]{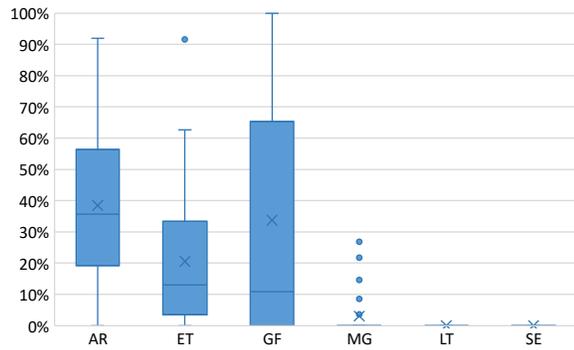}
    \caption{Summary of Detected Test Smells}
    \label{fig:group1}
    \vspace{-0.35cm}
\end{figure}

\noindent\textbf{Assertion Roulette(AR)}:
It is clear within~\autoref{fig:group1} that the AR smell is quite common within our set of VR projects. Indeed, 18\% of the projects we analyzed had assertion roulette within 20\% or more of their test cases.
% Furthermore, \texttt{watson-developer-cloud@unity-sdk} was the worst offender where this smell was found within 91.96\% of its test cases. This is especially concerning since this project is backed by IBM, a large company.

% \todo{Dhia: compress smell examples, use ... only keep what's necessary to understand the smell}
% FeedbackList listFeedbackResponse = null;
%         service.ListFeedback(
%             callback: (DetailedResponse<FeedbackList> response, IBMError error) =>
%             {    listFeedbackResponse = response.Result;
%   },
                % Assert.IsNotNull(listFeedbackResponse);
%             feedbackType: "element_classification",
%             includeTotal: true); 
        % Assert.IsNull(error);

\autoref{lst:assert_roulette_example} is an example of an assertion roulette from the \texttt{watson-developer-cloud@unity-sdk} project, where the test is attempting to verify whether the feedback object meets the developer's expectations. It would be difficult to diagnose the exact cause of this test's failures. For example, whether the feedback  response is null, or whether it's empty, as no messages are given in the assertion statements in lines 3 \& 4.

\lstset{style=mystyle}
\begin{lstlisting}[language={[Sharp]C}, caption={ Assertion Roulette Smell from \texttt{unity-sdk}}, label={lst:assert_roulette_example},basicstyle=\scriptsize]
[UnityTest, Order(7)]
public IEnumerator TestListFeedback(){...
    Assert.IsNotNull(listFeedbackRes.Feedback);
    Assert.IsTrue(listFeedbackRes.Feedback.Count>0);...}
\end{lstlisting}
\vspace{-0.26cm}

\noindent\textbf{General Fixture(GF)}:
The GF smell is quite widespread within our project set, as illustrated within~\autoref{fig:group1}, and was found in 18\% or more tests of 12 projects.
% However, we found 3 projects possessing this smell within 100\% of their test cases, \texttt{myieye@cube-arena}, \texttt{iamtomhewitt@vr-pacman}, and \texttt{vimeo@vimeo-unity-sdk}. 
% The latter of which is  especially concerning, since it's backed by Vimeo, a medium-sized company, and it is an SDK that meant to be used by other projects, and may cascade its problems within them.
%   EncoderManager encoder;
%   GameObject asset;
%  var holoplay = (GameObject)AssetDatabase.LoadAssetAtPath("Assets/HoloPlay/HoloPlay Capture.prefab", typeof(GameObject));
        % recorder.captureLookingGlassRT = false;
            % encoder.Init(recorder);
        % Assert.AreEqual(recorder.renderTextureTarget, null);
% public class EncoderManagerTest : TestConfig{
%     GameObject obj; VimeoRecorder recorder;
% holoplay, Vector3.zero, Quaternion.identity
%     ...

\lstset{style=mystyle}
\begin{lstlisting}[language={[Sharp]C}, caption={General Fixture Smell from \texttt{vimeo-unity-sdk}}, label={lst:gen_fix_example},basicstyle=\scriptsize]
GameObject obj; VimeoRecorder recorder;
[SetUp]
public void _Before() {...
    obj = new GameObject();
    recorder = obj.AddComponent<VimeoRecorder>();}
[Test]
public void Init_LookingGlass(){...
    asset = GameObject.Instantiate(...);
    Assert.AreEqual(recorder.renderTextureTarget,null);}
\end{lstlisting}
\vspace{-0.3cm}

\autoref{lst:gen_fix_example} represent a general fixture smell within \texttt{vimeo@vimeo-unity-sdk}, 
where the test is using the recorder  object initialized by the \texttt{\_Before()} fixture, but not the \texttt{obj} object. This causes a non-optimal memory consumption due to the initialization of an object that's not used by the test method. It may also make it more difficult to evolve the test cases since it may be difficult to determine which objects are being used by the tests. For example, the test case described above initializes and uses a new game object instead of the one initialized by the game fixture.

\noindent\textbf{Eager Test(ET)}:
For the ET smell, represented within~\autoref{fig:group1}, it is clear that this smell is also common within our VR project-set than the two previously-mentioned smells. Indeed, we found that 16 projects have this smell within 10\% or more of their test cases.
% \texttt{RussellXie7@Unity\_Hololens\_Dev} suffers most from this test smell category as well. 
% NormalizedCartesianCoordinates cartesian = new NormalizedCartesianCoordinates(Random.onUnitSphere);
% NormalizedCartesianCoordinates reconverted_cartesian;

% // Test cubemap
% CubeUVCoordinates cubemap = cartesian;
% reconverted_cartesian = cubemap;
% // Test octahedron map
% OctahedronUVCoordinates octahedron_map = cartesian;
% reconverted_cartesian = octahedron_map;\cartesian.data + " versus " + reconverted_cartesian.data
% cartesian.data + " versus " + reconverted_cartesian.data)

\lstset{style=mystyle}
\begin{lstlisting}[language={[Sharp]C}, caption={Eager Test Smell from \texttt{PlanetariaUnity}}, label={lst:eager_test_example},basicstyle=\scriptsize]
[Test]
public void CoordinateSystemsTest(){... 
    CubeUVCoordinates cubemap=cartesian;
    Assert.IsTrue(Miscellaneous.approximately(cartesian.data,cubemap.reconverted_cartesian.data),"...");
    OctahedronUVCoordinates octahedron_map=cartesian;
    Assert.IsTrue(Miscellaneous.approximately(cartesian.data,octahedron_map.reconverted_cartesian.data),"...");}
\end{lstlisting}
\vspace{-0.3cm}

\autoref{lst:eager_test_example} is an example of an eager test from \texttt{mattrdowney@PlanetariaUnity}, where the test is attempting to verify that the cubemap and octahedron maps are storing and converting coordinates correctly from their Cartesian form. However, it would be difficult to diagnose which of these coordinate conversion process is failing, since the test would fail if either or both are problematic. 

\noindent\textbf{Mystery Guest(MG)}:
Unlike the aforementioned test smells, Mystery Guest is less common within our project set, as shown by the results within~\autoref{fig:group1}. Indeed, only 3 projects had this smell within 14\% or more of their test cases. 
% \texttt{Willychang21@MapboxARGame} had the most test cases of this smell with a rate of 26.8\%. 
% private const string _dbName = "UNITTEST.db";
% 	string[] tilesetNames = new string[] { TS_CONCURRENT1, TS_CONCURRENT2, TS_CONCURRENT3, TS_CONCURRENT4 };
% 	foreach (string tilesetName in tilesetNames)
% 	{
% [OneTimeSetUp]
% public void Init(){
% 	_cache = new SQLiteCache(_maxCacheTileCount, _dbName);
% }
% [Test]
% public void VerifyTilesFromConcurrentInsert()
% {
% "tileset '{0}' does not contain expected number of tiles", tilesetName
% \public class EncoderManagerTest : TestConfig{
    % ...
    
\autoref{lst:myst_example} represents a mystery guest smell within \texttt{Willychang21@MapboxARGame}. The issue this smell causes is that this test will fail if it is run within an environment where the database is not available, such as a CI environment or a different machine. This smell goes against the recommendation of using a mock object to represent an external resource programmatically during testing. 
\lstset{style=mystyle}
\begin{lstlisting}[language={[Sharp]C}, caption={Mystery Guest Smell from \texttt{{MapboxARGame}}}, label={lst:myst_example},basicstyle=\scriptsize]
private SQLiteCache _cache;
[Test]
public void VerifyTilesFromConcurrentInsert(){
    Assert.AreEqual(_tileIds.Count,_cache.TileCount(tilesetName),"...");} 
\end{lstlisting}
\vspace{-0.28cm}

% This project is a toolbox for map generation and thus needs access to data, which is saved within different files and SQL databases that this project uses within some of its test cases. 
\noindent\textbf{Lazy Test (LT) and Sensitive Equality(SE)}:
We did not discover any examples of these smells within the manual inspection used to verify the accuracy of our smell-detecting tool.
% We also verified our tool's capability to detect these smells with manually developed examples, as discussed within~\autoref{sub:sub:sec:eval_auto}. 
% Hence, we trust the automatic analysis result that  did not find any instances of these smells in our project-set. 
In the case of the SE smell, we found that most comparisons of objects within test cases were based on a specific object's property. Since most objects did not override the default \texttt{ToString} method, it makes the evaluation for comparative purposes based on their textual representation of limited usefulness. 
Furthermore, based on the findings in~\autoref{sub:sec:test_case_extent}, it's clear that the developers of the VR project-set write much less test code than functional code, 
%in comparison to the general recommendation~\cite{loc_ratio} and other C\# projects~\cite{Williams_Kudrjavets_etal_2009}, 
thus making it less likely that they would write multiple tests for the same method. 
In fact, the results we found regarding the ET smell earlier within this section point to the opposite practice of testing multiple functional code methods within the same test being the more common, if ill-advised, practice. 

% Contextualizing the results within those of the test quality of open-source Android apps as studied by Peruma et al.~\cite{Peruma2019}, while 58.46\% of files analyzed, an average of 38.43\% of test within VR projects we analyzed possessed this smell. A similar trend is noted for the other smells as well, where Eager Test was found in 38.68\% of files, it was found on average within 20.52\% of a VR project's test, as well as for Lazy test , sensitive equality and mystery guest smells, where the means were close to 0\% for VR projects, and respectively were 29.50\%,9.19\% and 11.65\% of Android projects' files. The only exception is General Fixture, found within 11.67\% of Android projects files', it was found on average within an average of 33.72\% of a VR projects' tests. 

We contextualized the overall test smell detection results from VR projects by comparing them with those from a similar test smell study of open-source Android applications by Peruma et al.~\cite{Peruma2019} 
% and briefly compare the test quality between these two types of applications.
While 58.46\% of files analyzed in Android exhibited the Assertion Roulette smell, an average of 38.43\% of tests within VR projects we analyzed possessed this smell. 
A similar trend is noted for the other smells as well, where Eager Test was found in 38.68\% of Android projects' tests, it was found on average within 20.52\% of our VR projects' tests. For Lazy Test, Sensitive Equality, and Mystery Guest smells, the three means were close to 0\% for VR projects, and respectively were 29.50\%, 9.19\%, and 11.65\% within Android projects. The only exception is General Fixture, which was found within 11.67\% of Android projects' tests and within an average of 33.72\% of VR projects' tests. Overall, both application types show a comparable and problematic frequency of the different test smells. However, the situation in VR applications is considerably worse considering the lack of prevalence and effectiveness of its testing overall.

% The tests smells presence within the Android applications which we used as a baseline is considered substantial~\cite{Peruma2019}, thus making their comparable presense in our set of VR applications substantial as well, establishing that they too suffer from important and prevalent test design problems. 
% \todo{Highlight VR specific categories in taxonomy}
% \todo{re-organize taxonomy to be closer to unity organization}
% \todo{add definitions for technical terms such as prefab, AssetBundle,... }
\vspace{-0.23cm}
\subsection{What are the different categories of VR Test Cases and which categories are specific to VR?}
\label{sub:sec:taxonomy}
\begin{figure*}
    \centering
     \includegraphics[width=0.98\linewidth]{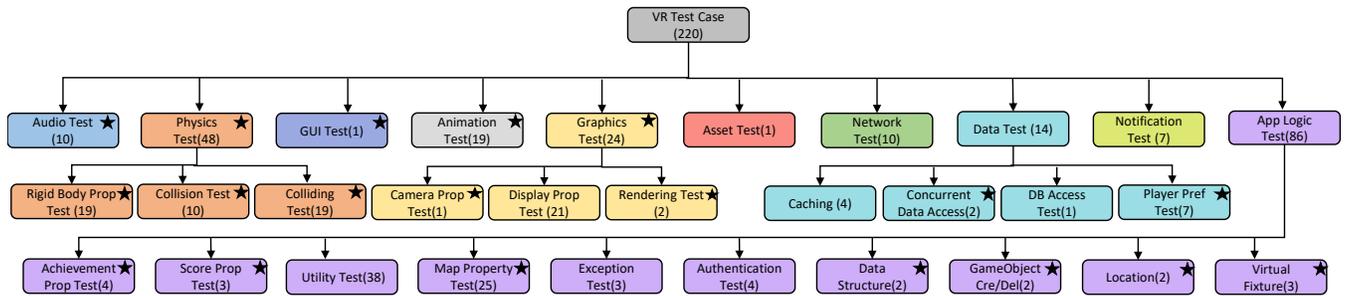}
      \vspace{-0.1cm}
    \caption{Taxonomy of VR Testing Categories ($\filledstar$ indicates a VR-specific category)}
    \vspace{-0.6cm}
    \label{fig:rq4}
\end{figure*}

\vspace{-0.15cm}
%\todo{Nafees}
To understand the test types that reflect the characteristics of VR applications, we followed the methodology discussed in Section~\ref{sub:sec:method_manual_analysis} and carried out a manual analysis of 220 test methods from four VR projects. 
% These projects have been selected from all projects with executable test methods that reached a code coverage of at least 10\%.
We divided all the test methods into ten different main categories. ~\autoref{fig:rq4} represents the taxonomy that we have generated based on a parent-child hierarchy. The digits indicate the number of test cases we discovered for the corresponding category. In addition, we conducted a second manual analysis on the test cases of non-VR Unity Projects to identify the test categories that are unique to VR projects, indicated by stars within~\autoref{fig:rq4}. In the rest of this section, we will carefully discuss each of the test categories as well as their sub-categories by giving detailed definitions. 
% And code examples. And we will specifically focus on the VR nature related categories.
We present code examples to help avoid the bias of definition descriptions, but these definitions can also be expanded to general cases in other VR frameworks.

% As mentioned in section 3.3, we assessed 220 distinct test cases from several VR unity projects and classified them appropriately for manual analysis. Due to a lack of resources and automated tool assistance, this categorization is an extremely time-consuming human process.
% \\Based on their characteristics, we organized diverse test methods into a hierarchy-based framework. We divided the test methodologies into ten distinct VR-related categories. The overall number of occurrences of these categories in the examined test methods was also highlighted. We sub-categorized a few of them based on the primary category to make it more detailed. Figure 6 represents the taxonomy that we have generated based on parent-child hierarchy. The following are the primary distinct categories that we considered:

% \textbf{1. Audio Testing}:
% To play sounds in a 3-D world, an Audio source is attached to GameObject. This test is related to unity Audio API like AudioSource.Play, AudioSource.Pause, AudioSource.Stop.

\subsubsection{Audio Test}
Audio Test is the test that validates if the sound works correctly, such as playing, pausing, and stopping the sound clips.
The sound design of VR projects is supposed to provide users with a believable experience.
Rather than being separate from the environment and only coming from a single direction, sounds in a virtual world are designed to be more interactive.
% They can come from behind or above users based on their positions and interactions with the environment.
% For example, users may hear sounds behind them after triggering an achievement by moving forward. In addition, users can also sense a sound coming toward them when approaching some specific scenes.
We found 10 test methods that fall into this category.
% and the Unity audio APIs we identified are \texttt{AudioSource.Play}, \texttt{AudioSource.Pause}, and \texttt{AudioSource.Stop}.

\subsubsection{Physics Test}
Physics Test ensures the physical interaction between objects is correctly represented in effects like collision, falling due to gravity, and other forces. The key concept of physics is to have one or more forces that apply to objects. 
% In the following, we will discuss each of the
Its subcategories are:

\noindent\textit{\textbf{Rigidbody Property Test}}. 
Rigidbody is a component of GameObject that enables the Unity physics engine control. It allows interaction with real-time physics, which includes forces, gravity, mass, and momentum. The Rigidbody Property Test is primarily concerned with physical behaviors such as movement and rotation.
We identified 19 of them in total.

\noindent\textit{\textbf{Colliding Test}}. 
Colliding in VR represents an interaction between two or more objects that does not trigger the physical collision effect.
Colliding can be divided into three stages: at the start of colliding, during colliding, and after colliding. 
Colliding Tests evaluate the correctness of this dynamic process. 
% Unity provides a group of APIs to manage the colliding: \texttt{OnTriggerEnter}, \texttt{OnTriggerStay}, and  \texttt{OnTriggerExit}.
% These methods are invoked when physics engine detect the overlapping of more than two objects' Colliders.
% Unity defines a Collider as the boundary of the object. 
% First, \texttt{OnTriggerEnter} is invoked when a contact between two colliders happens,
% then, \texttt{OnTriggerStay} is invoked to update every frame during the contact. 
% eventually, \texttt{OnTriggerExit} is called when contacting ends and colliders no longer overlaps.

\lstset{style=mystyle}
\begin{lstlisting}[language={[Sharp]C}, caption={Colliding Test from \texttt{vr-pacman}}, label={lst:colliding},basicstyle=\scriptsize]  
[UnityTest]
public IEnumerator CollidingWithTeleporterMovesPlayer(){
    Vector3 positionBeforeTeleport = pacman.transform.position;
    pacman.transform.position = teleporter.transform.position;
    yield return new WaitForSeconds(WAIT_TIME);
    Assert.AreNotEqual(positionBeforeTeleport, pacman.transform.position);}}
\end{lstlisting}
\vspace{-0.25cm}

The test design we observed in the test cases follows the three previously-described stages. The test method will first record tested properties and targets before colliding. Then, it will trigger the colliding condition by updating the interaction between the objects. After waiting for a short system response time, the test method will collect the new value of the tested properties and targets. Eventually, an assertion will be carried out between old values and new values or between new values and expected values.

Listing~\ref{lst:colliding} shows an example of a colliding test from the project \texttt{iamtomhewitt@vr-pacman} in the file PacmanCollisionTests.cs.
%The test case first set speed value to \texttt{50f} for PlayerControl object \texttt{pcn} in line 3.
%Then it makes \texttt{player} position overlap with \texttt{sup} position in line 5, where \texttt{sup} is a GameObject with speeding up tag.
%This overlapping will then trigger the colliding event, which results in the corresponding speed value changes.
%Specifically, \texttt{OnTriggerEnter} method is called, and an extra value of \texttt{20f} has been added to the speed value.
%At line 7, the test case inserts an assertion to check if the new speed value has been updated after a constant waiting time.
The test case first sets the initial position for GameObject \texttt{pacman} in line 3.
Then, it makes \texttt{pacman} position overlap with the \texttt{teleporter} position in line 4 by assigning the same value to these two positions.
The overlapping of two colliders then triggers the colliding handling event, which results in \texttt{pacman} position updating.
Specifically, \texttt{OnTriggerEnter} method is called by Unity, and the corresponding x, y, and z values of \texttt{pacman} are updated by the value of \texttt{teleporter}.
After a waiting time, an assertion is inserted at line 6 to check whether the new position is different from the initial position.
In total, we identified 19 colliding test methods.

\noindent\textit{\textbf{Collision Test}}. 
Collision is very similar to colliding in VR. The major difference is that a collision will trigger a physical collision effect.
For example, when a car is driving toward a wall, colliding will let the vehicle pass through the wall without stopping, but a collision will result in a crash where the vehicle hits the wall and stops.
Collision Tests verify the correctness of the expected functionalities in regards to the process of collision. 
Similar to colliding, collision also includes three stages: entering, during, and after the collision. We found 10 tests of this category.
% Unity provides another set of methods to handle the collision events: \texttt{OnCollisionEnter}, \texttt{OnCollisionStay}, and  \texttt{OnCollisionExit}.
% The test design in collision also follow the similar pattern in colliding.

%Listing~\ref{lst:collision} shows an example of a collision test from the project \texttt{iamtomhewitt@jet-dash-vr} in file PlayerCollisionTests.cs.
% In line 3, both \texttt{player} and \texttt{stationaryObstacle} are GameObjects that have been instantiated in \texttt{Setup()}.
% By default, the player object is moving forward and not dead. 
% The test method triggers the collision intentionally by making the two objects' positions overlap in line 3.
% The Unity physics system detects this collision by monitoring the colliders of thess two objects and then invokes the \texttt{OnCollisionEnter} API to play the animation of the collision effect, stop the player's object and mark it as ``dead''.
% Finally, An assertion statement in line 6 is to check if the value of the dead property is true.
% In total, we identified 10 collision test methods.

% \lstset{style=mystyle}
% \begin{lstlisting}[language={[Sharp]C}, caption={Collision Test in \texttt{iamtomhewitt@jet-dash-vr}}, label={lst:collision},basicstyle=\scriptsize]  
% [UnityTest]
% public IEnumerator ShouldDieWhenCollidingWithStationaryObstacle() {
%     player.transform.position = stationaryObstacle.transform.position;
%     yield return new WaitForSeconds(TestConstants.WAIT_TIME);
%     Assert.IsTrue(pc.IsDead());
% }
% \end{lstlisting}

\subsubsection{GUI Test}
GUI Tests make sure that visualized interface components such as text, layouts, and input fields in VR applications work as specified. 
In traditional software, all the visualized items can be categorized as GUI components.
In VR applications, both the GUI and the scene can be visualized at the same time. The scene may be changed by the camera position but the GUI may remain the same.
% Unity defines classes such as \texttt{Canvas} to hold all GUI elements inside.
The one test example we found evaluates the conversion from numeric values to GUI text.
\subsubsection{Animation Test} 

Animation Tests verify the correctness of the expected functionalities and properties when target objects are in action. 
For example, location updates, moving, stopping, and rotation.
The test cases we observed all follow a time-order-oriented design, where  an assertion is inserted to compare the status before and after time frame updates. Unity provides a PlayMode testing scheme that executes the test code dynamically when the VR application is running.
% The Unity framework divides
% the testing scenarios into two types (motion and motionless)
% and provides two corresponding test modes, EditMode and
% PlayMode. 
% The test methods in EditMode mode are very similar to those of traditional software, and they usually do not update based on time. 
% However, test methods in PlayMode take the time dimension into consideration and represents the motion in VR by frame updates.
% This principle is very similar to how video is played, which shows multiple distinct still image frames in one second.
% Unity provides APIs like \texttt{Update}, \texttt{FixedUpdate}, \texttt{Awake}, and \texttt{Start} to support the frame or time based updates of an object. 
% The test design we observed follows a time-oriented order, and eventually, an assertion that compares the status before and after the frame update is inserted.
% One of the most significant terms in virtual reality is motion property. Animation States in the Mecanim StateMachines use motions. It refers to an object's modified position, speed, and location after its initial value has been changed.
\lstset{style=mystyle}
\begin{lstlisting}[language={[Sharp]C}, caption={Animation Test from \texttt{vr-pacman}}, label={lst:motion},basicstyle=\scriptsize]  
[UnityTest]
public IEnumerator ResetPositionsWork(){
    Vector3 pacmanPos = pacman.transform.position;
    Vector3 ghostPos = ghost.transform.position;
    goManager.StartMovingEntities();
    yield return new WaitForSeconds(WAIT_TIME);
    goManager.StopMovingEntities();
    goManager.ResetEntityPositions();
    Assert.AreEqual(pacmanPos,pacman.transform.position);
    Assert.AreEqual(ghostPos,ghost.transform.position);}
\end{lstlisting}
\vspace{-0.25cm}

Listing~\ref{lst:motion} shows an example of an Animation Test from the project \texttt{iamtomhewitt@vr-pacman} in GameObjectManagerTest.cs.
In lines 3 and 4, two GameObjects' positions have been recorded into two 3-D vector objects.
From lines 5 to 8, the \texttt{goManager} object commands all the objects in the scene to move, stop and then reset.
Then the \texttt{Update} APIs defined in all GameObjects will take the corresponding actions to update their positions frame by frame.
Eventually, in lines 9 and 10, the test oracles are inserted to compare the positions value update.
In total, we identified 19 Animation test methods.

\subsubsection{Graphics Test} 
Graphics Test tests the appearance of VR application. It includes Camera, Rendering, Display, Lighting, etc.
The following subcategories are based on the test cases observed.

\noindent\textit{\textbf{Camera Property Test}}. 
In VR applications, developers design one or more cameras in the virtual environment to provide users with different aspects of an immersive experience.
Depending on the viewport space point, which is a relative point to the camera (i.e., the field of view based on camera features like locations, rotations, and scale), objects will be rendered in the VR world by following different performance-optimization purposes.
Based on these special features, Camera Property Tests involve camera functionalities, such as camera counting, camera moving, rendered objects checking, etc. 
We identified one method in total.

Listing~\ref{lst:camera} shows an example of Camera Property Test from project \texttt{iamtomhewitt@jet-dash-vr} in the file PlayerControlTests.cs. %In Unity, an unlimited number of cameras can be used to render a scene. 
This project has two preset cameras in the scene, VR Camera and a Normal Camera.
In line 3, GameSettingsManager \texttt{gs} activates its VrMode by passing a boolean value true to \texttt{SetVrMode}.
Next in line 4, PlayControl \texttt{pc} initializes the VR scene by invoking the \texttt{Start} method, where the VR viewpoint will be provided to users, as well as the corresponding audio listener and clip management.
The final assertion in line 5 is to examine the existence of a Normal Camera with the tag ``MainCamera''.

\lstset{style=mystyle}
\begin{lstlisting}[language={[Sharp]C}, caption={Camera Property Test from \texttt{jet-dash-vr}}, label={lst:camera},basicstyle=\scriptsize]  
[UnityTest]
public IEnumerator ShouldReadGameManagerSettingsCorrectly(){
    gs.SetVrMode(true);
    pc.Start();
    yield return new WaitForSeconds(TestConstants.WAIT_TIME);
    Assert.AreEqual(GameObject.FindGameObjectWithTag("MainCamera"), null);}
\end{lstlisting}
\vspace{-0.25cm}

\noindent\textit{\textbf{Display Property Test}}.
The display property test focuses on checking if objects are correctly displayed in VR scenes after rendering. Developers usually apply updates to the current field of view, then insert an assertion to check the results.
Listing~\ref{lst:display} shows an example of a display property test from the file MapboxUnitTests\_TileCover.cs in the project \texttt{willychang21@MapboxARGame}.
The project is an SDK for 3-D map generation, and this test method checks the correctness of tile updates when zooming. Tiles in this context are basic representations of parts of the map.
From line 3 to line 5, the variable \texttt{zoom} increases from 0 to 7 to represent the different levels of zoom.
In line 4, the test method updates the scene, \texttt{Vector2dBounds.World()}, using the \texttt{zoom} value. The \texttt{Get()} method performs the mathematical operations and returns the updated value of \texttt{tiles}. The assertion checks the display by confirming the value of the tile count. 21 tests of this category were found.
\lstset{style=mystyle}
\begin{lstlisting}[language={[Sharp]C}, caption={Display Property Test from \texttt{MapboxARGame}}, label={lst:display},basicstyle=\scriptsize]
[Test]
public void World() {
    for(int zoom=0; zoom<8; ++zoom){
        var tiles=TileCover.Get(Vector2dBounds.World(),zoom);
        Assert.AreEqual(Math.Pow(4,zoom),tiles.Count);}}
\end{lstlisting}
\vspace{-0.25cm}

% Listing 6 shows a test that is related to display test. If zoom level is greater 8 then it would generate so many tiles that would make memory shortage. On a map, zoom level determines the area that can be seen. Tiles are the smallest building blocks for an image. As part of the test, they are checking the tile count to ensure a proper display area is being displayed.

\noindent\textit{\textbf{Rendering Test}}.
A VR world consists of multiple visible and invisible entities. 
The goal of rendering is to create these entities in a user's field of view (camera view).
To obtain a more realistic result, developers need to consider factors like texturing, lighting, image effects, etc. 
A rendering test checks the correctness of these processes.
For example, in a VR map-based project, one of the tests may check if the tiles have been successfully loaded onto the scene. We found 2 tests belonging of this category.

\subsubsection{Asset Test} 
An asset is any item that the developer uses to create the VR application. 
Assets include visual, audio, or  other elements like models and textures. 
An asset can either be from outside of Unity as a file or inside of it as internal data from the editor.
Asset Tests verify the assets during the importing, loading, unloading, distributing, and other processes. 
Correctly managing a large number of assets automatically is a demanding task for VR developers. We found 1 test of this category.

\subsubsection{Network Test}
Networking and multiplayer support are two features in VR applications that require local or wide-area network access.
Network tests check whether network calls and responses are correct, such as response count and response data. 
They ensures that the receiving end is configured correctly and that data is transmitted between the sender and the receiver without any issues.
Testing a real-time multiplayer feature ensures that the connections between clients, servers, and hosts are working correctly. 10 Tests of this category were found.
% Network Test checks whether network calls and responses are correctly addressed, such as response count and response data. 
% The Real-time Multiplayer feature is also widely seen in VR applications.
% This test ensures that the receiver end is configured correctly and that data is transmitted between the sender and receiver without interruption.

\subsubsection{Data Test}
Data tests involve dataset evaluation through caching systems, local file systems, or other local and remote databases.
These tests confirm that a VR application is storing, retrieving, deleting, and updating data appropriately. For example, testing whether multiple objects can be added to a cache system concurrently. 
Its subcategories are Caching Test, Concurrent Data Access Test, DB Access Test, and Player Pref Test.
PlayerPref is a built-in Unity API that helps developers quickly access internal data between frames and across multiple VR scenes.
For example, PlayerPref can store a Player's personal preferences and be loaded between different VR sections to provide a consistent user experience. 
In total, we identified 14 Data Tests. 
% directories and  and that there is enough storage so that the program does not terminate unexpectedly due to a lack of space.
% Data storage testing involves testing different properties such as caching, concurrent data access, and database access.

% \noindent\textit{\textbf{Player Pref Test}}.
% PlayerPrefs are Unity built-in APIs that allow storing the Player's preferences.
% It helps developers easily access global data between frames and across multiple VR scenes.
% The seven test methods we observed mainly focus on checking the preference data accessing and manipulating through functions like \texttt{PlayerPrefs.SetInt} and \texttt{PlayerPrefs.GetInt}.

% PlayerPrefs is used to store preferences meanwhile it also provides modest storage so that we can save data without having to access local data file systems. This test covers two key PlayerPrefs functions: PlayerPrefs. SetInt() and PlayerPrefs. GetInt().

\subsubsection{Notification Test} 
Notification systems are widely used in traditional and mobile software to deliver messages to recipients.
So far, there are no built-in notification APIs provided by Unity for VR applications. Developers customize their own notification system and may present messages in a variety of formats.
Notification Tests ensure that the notification system behaves as expected. Based on our observations, no matter what the format of the notifications is, they are tested in PlayMode.
For example, when certain conditions are dynamically triggered, developers pause the updates for all other GameObjects by invoking \texttt{StopAllCoroutines()} API. Then, they only focus on the animation of the notification message delivery. Eventually, developers resume  updates by calling \texttt{StartCoroutine()} for the other objects. We found 7 tests of this category.

% Notification systems in VR are similar to normal software. One major difference is that notification tests are PlayMode tests.
% Based on our observation, when certain conditions are triggered dynamically, developers pause the frame updates for all other GameObjects by invoking \texttt{StopAllCoroutines()} API. Then the following frames will only play the animation of the notification. After the latter stops, developers resume the frame updates by calling \texttt{StartCoroutine()}.
% Notification message generally pops up when an object reaches a  predetermined value.
% One of the most crucial aspects of virtual reality games is to reward players and provide a sense of progression. If the pacman eats a ghost in the VR Pacman game, for example, it should be rewarded with bonus points.

% \subsubsection{Asset Test} 
% As part of the VR application scene, the application can load objects (e.g., Asset, Prefabs, Subtitles, etc.) from  AssetBundle to ScriptableObject. The application can make run-time modifications to those objects and save those modifications back to AssetBundle. Asset Test involves loading and unloading is to verifying whether Assets, Prefabs and other objects are loading correctly and can be modified and unloaded to AssetBundles to save the modifications properly.

\subsubsection{App Logic Test} App Logic Test is relevant to a specific VR application based on its business logic. 86 tests from this category were found. To provide a more complete understanding of the App Logic Test category, we divided it into ten different sub-categories. (1) Achievement Property Test:  evaluates the correctness of the achievement system
% , for example, testing whether the awards and records can be successfully tracked and unlocked.
(2) Score Property Test: works closely with the achievement system in VR Games to keep track of a player's progress
and give them rewards or penalties based on their performance.
(3) Utility Test: a collection of criteria tests that evaluate utility-related methods, which are methods that execute common, often-repeated functions. (4) Map Property Test: verify that  vector data files are correctly displaying the map. It can be generalized to evaluate if a VR scene is displayed properly.
% The features which concern these tests are boundaries, zoom levels, tracepoints, etc.
(5) Exception Test: check if a test method behaves as intended during the execution of the program. (6) Authentication Test: ensure a correct data verification using token error messages and token statuses. (7) Data Structure Test: verify the implementation correctness of customized data structures. (8) GameObject Creation and Deletion Test: evaluate the dynamic creation and deletion of GameObjects within PlayMode.
% For example, it tests if a GameObject is successfully dynamically instantiated from loading a resource file. The assertions are check unique features of the generated object, such as its tag information.
(9) Location Tests: verify the position values (e.g., longitude, latitude) in a 3-D world. Finally, (10) Virtual Fixture Tests: check the correctness of the presentation of a pre-defined navigation path in a virtual environment.  

% that is  to check the particular feature of the  and of this test is the creation of GameObjects from a particular directory using prefab. They deleted GameObjects after certain operations were applied to them to improve memory management. 

% They initially fixed the position for pacman and ghost in listing 5 which is an example of display test from project \texttt{willychang21@MapboxARGame}. After that, StartMovingEntities() and StopMovingEntities() are called to start and stop the objects, respectively, before returning to their original position.
% Finally, they compared the latest position of those items to the initial position in the test case.

% \todo{discuss: put achievement and scoring to game property test, put Text Object Property to utility test}

We believe this taxonomy allows VR developers to identify the different areas and aspects of testing they should focus on within their projects. This is especially important due to the apparent lack of established VR testing practices, as well as the several issues noted with testing prevalence, effectiveness, and quality, as found within ~\cref{sub:sec:test_case_extent,sub:sec:test_case_effectiv,sub:sec:test_case_quality}, which is concerning since modern VR headsets and their applications have been around since 2016~\cite{Virtual_reality_2022}. 
% \todo{add discussion between the VR and non-VR project}
% address Reviewer1-7
% explain the reason for this comparison
% Unity platform not just supports VR application development but also supports non-VR applications. %such as 2D games, asset management tools, etc. 

It's important to note that the Unity platform supports not only support the development  of VR applications but also that of non-VR applications. Thus, various Unity APIs can serve similar functionalities in both types.
Consequentially, the taxonomy we defined may also partially apply to test cases in non-VR applications. It is difficult to clarify which APIs are specific to VR by only studying the official documentation. To identify the test categories unique to Unity-based VR projects, we constructed a set of non-VR projects through a process detailed within~\autoref{sub:sec:dataset}, and conducted a second manual analysis on their test cases to determine which test categories they did not cover.
% explain the dataset and what we have found
% In particular, we follow the same data collection standard and get four non-VR projects with at least 100 commits and over 10\% test coverage. 
% Of the four non-VR projects, two are 2D game applications, and the other two are tools for asset management and texture generation in Unity.We labeled their 281 test cases and found that four major categories and four subcategories of two major categories from the taxonomy were not present within them.
% in any of the test cases in these non-VR projects.
The four missing main categories were Audio Test, Physic Test, GUI Test, and Animation Test. The four missing sub-categories were Camera Prop and Rendering Test from Graphic Test, Player Pref Test, and Concurrent Data Access from Data Test.
%detected test taxonomy: App Logic Test, Asset Test, Network Test, Caching Test, Authentication Test, DB Access Test, Notification Test, Display Test, Exception Test
%non-detected test taxonomy: Audio Test, Physic Test, GUI Test, Animation Test, Graphic Test (Camera Prop, Rendering Test), Data Test (Player Pref)

% discuss the potential reasons
% Next, we will briefly discuss our findings about these non-exist taxonomies.
Based on the previously-detailed definitions, Physics Test, Animation Test, Camera Prop Test, Rendering Test, and Player Pref Test evaluate the correctness of the simulated world's features. GUI Test, Audio Test, and Concurrent Data Access Test vary depending on a project's implementation and design.
Two of the non-VR projects analyzed are non-graphical libraries, also referred to as Tools, and do not simulate virtual environments. Hence, they contain none of these test categories.
The other two are respectively a 2-D card Game and a 3-D puzzle Game. Both present a single view of the entire virtual world and are not designed to dynamically update a user's viewpoint and regenerate the graphics. Hence, it's not surprising that they contain neither Camera Prop Tests nor Rendering Tests.
Also, no Physics Tests were found because these two projects do not design the physics effects between their objects. In other words, all objects do not move and will not overlap with each other.
% within them as they do not handle the physics system effects.
Although we did not identify any test cases in the GUI Test, Audio Test, and Animation Test in the two non-VR Games, we detected their related APIs in their source code. These findings hint at a lack of testing in current non-VR Game applications similar to their VR counterparts.
% \vspace{-0.3cm}
\section{Threats to Validity}
\vspace{-0.15cm}
\label{sec:discussion}
%%\todo{Xue}
On construct validity, the main threat is the soundness of the automatic analysis results.
To measure the design quality of the test cases, we followed the work of De Bleser et al. ~\cite{de_bleser_2019} and selected six test smell types, including Assertion Roulette, General Fixture, Sensitive Equality, Eager Test, Lazy Test, and Mystery Guest. To further validate the automated reports, we manually labeled smell types of 220 test methods, with a 0.92 kappa score on average, among the two co-authors who worked separately. Automatic test smell detection results show an average of 91.35\% accuracy score, a 92.62\% recall, and a 91.98\% F-1 score. Hence, we believe our automatic analysis results correctly represent the quality of VR test cases.

On internal validity, the threat is  potential bias when answering RQ4.
Since there is no existing taxonomy to use as a reference, three co-authors reviewed and categorized VR test methods separately. Then, they carried out voting and discussion to finalize the results.
In addition, when defining the VR test taxonomy, the paper uses a general language and adds code examples to reduce the gap between the theory and the observed practice. 

On external validity, the main concern is the representativeness of studied projects.
% As a matter of fact, it is nearly impossible to cover all types of VR apps with the current fast growing speed of the industry and new VR apps being released every month.
Our empirical study consisted of 97 open-source Unity VR projects from Nusrat et al.~\cite{Nusrat_2021}.
These projects have been carefully selected with at least 100 commits per-project, and the dataset covers 7 different versions of Unity frameworks (from Unity 4 to Unity 2021). Furthermore, we considered both small-scale projects and large  projects backed by companies. 
% TODO: add number of industrial vs indie projects. 
This allows us to uncover insight that may reflect general characteristics of all VR applications.
Moreover, the taxonomies defined in RQ4 do not depend on Unity alone and can be extended to more general scenarios. In addition, the 4 non-VR projects used within the analysis to detect the VR-specific test categories were selected using the exact same criteria as their VR counterparts, and contain both Tool and Game projects. Thus, we believe our findings provide insight for future test automation design. 

\section{Related Works}
\vspace{-0.11cm}
\label{sec:related}
%%\todo{Foyzul}
\subsection{Study on Software Testing}
\vspace{-0.19cm}
Software testing is an essential but costly and effort-intensive activity of software development. This has pushed the research community to study software testing practices. Greiler et al.~\cite{Greiler2012} conducted a qualitative study in which they interviewed 25 practitioners about how they test Eclipse plugins. They identified that unit testing plays a critical role, whereas integration problems are identified by the community. Kochhar et al.~\cite{Kochhar2013} performed a study on 20,000 non-trivial software projects and explored the correlation of test cases considering various factors. The study discovered that as projects grow in size, their ratio of test cases per LOC decreases. %As the number of developers grows, the number of test cases per developer decreases.
% Memon et al.~\cite{Memon2017} analyzed continuous testing practices at Google, and identified that certain frequently modified code structures cause more breakages than others, and that code recently modified by multiple developers breaks more often. Similar to that work,
Pecorelli et al.~\cite{Pecorelli2020TestingOM} performed an empirical study targeting  1,780 open-source Android apps and identified that the effectiveness of their test cases is low and  that they suffer from quality issues. %For accessing test quality, Peruma et al.~\cite{Peruma2019} performed a study on 656 open-source Android applications. The study found a widespread occurrence of test smells in apps and apps tend to have test smells early in their lifetime. % with different degrees of co-occurrences with different test smell categories. 
% More recently, Wang et al.~\cite{Wang2021} performed an empirical study on test adoption practices and test case generation tool-support in the context of Machine Learning (ML) applications. %They discovered that most of the ML libraries do not maintain a high-quality unit test and there is limited support for automated unit test case generation of ML libraries.
Several other studies~\cite{Bowes2017,Palomba2017OnTD,Palomba2017DoesRO,Nejadgholi2019,cadar2011symbolic} also performed empirical analyses of software testing practices and different aspects of testing adoption. Even though these works performed empirical analysis on general-purpose and Android app testing practices, none of the work performed an analysis on testing practices of VR applications.
\vspace{-0.25cm}
\subsection{Study on Game Development and VR}
\vspace{-0.19cm}
As Game and Virtual Reality (VR) applications are becoming more popular and accessible, the research community has started studying the development practices of Game and VR projects. To identify common bugs in Game applications, Truelove et al.~\cite{Truelove2021} performed empirical analysis on 12,122 bug fixes from 723 updates of 30 popular games. 
% Pascarella et al.~\cite{Pascarella2018} investigated how developers contribute to video game projects vs. non-game projects by analyzing different kinds of artifacts within these projects, how their developers handle malfunctions, and how they perceive the development process. They identified that developing games requires more diverse resources than developing non-game applications, and that it's difficult to reuse Game application code.
% Wu et al.~\cite{Wu2020ICMSE} proposed a machine learning-based approach to identify regression tests for Game applications.
Politowski et al.~\cite{Politowski2021ASO} performed a survey to understand existing testing practices within Game development. Recently Nusrat et al.~\cite{Nusrat_2021} performed a study on 100 Unity VR applications and created a performance bugs taxonomy in the context of VR. 
% Borrelli et al.~\cite{BorrelliACM2020} proposed the static analysis tool UnityLinter to identify code smells in Unity-based projects. 
% Recently, Ashtari et al.~\cite{Ashtari2020} interviewed 21 AR/VR creators from different groups such as hobbyists, domain experts, and professional designers, and concluded pointed out that testing VR applications is a complicated task.
Although there are several works related to VR and Game development, none of them analyzed VR-testing's adoption, practices, and characteristics by analyzing existing code bases. In this work, we tried to fill this knowledge gap by analyzing existing projects and their testing practices.
% \vspace{-0.2cm}
\section{Implications}
\label{sec:implications}
\vspace{-0.15cm}
%%\todo{Foyzul}
Our analysis pointed out some critical factors and observations for VR application testing which can be beneficial for VR developers, tool builders, and researchers. The findings of this paper lead to the following implications.

\noindent\textbf{For VR Developers:}
\texttt{RQ1}, \texttt{RQ2}, and \texttt{RQ3}  clearly point out that VR testing's effort and effectiveness are low and that it suffers from quality issues. 
% As testing is one of the most critical quality aspects of any software application, VR developers and testers should put more effort into improving quality of tests for VR applications. At the same time,
% Via \texttt{RQ3}, we identified that VR test cases suffer from quality issues 
% such as Assert Roulette, Eager Test, etc. 
% Our aim with  research was to
These results  illustrate the problematic state of testing within VR applications, and that VR developers and testers should put more effort into improving the quantity, coverage, and quality of their tests.
% especially its prevalence, quality, and effectiveness,
% and can help guide future research that can identify the causes behind the current state of testing.
% However, through our findings, it's clear that VR developers should pay extra attention to avoid quality issues while developing test cases.
In addition, \texttt{RQ4} generates a VR test case taxonomy based on testing goals, which can be helpful for VR developers as a general guideline that directs the formulation of  their test cases.

\noindent\textbf{For VR Tool Builders:}
\texttt{RQ2} points out the necessity of tool support for VR application code coverage analysis. Even though Unity provides tool support for code coverage measurement, the tool is unusable in a lot of cases due to compatibility issues. Similarly, based on findings from \texttt{RQ3}, tool developers should develop tool support for automatic detection and repair of test smells for VR applications.

 \noindent\textbf{For Software Engineering Researchers:}
 From \texttt{RQ1} \& \texttt{RQ2}, it is evident that  VR application testing is not sufficient. Researchers can do further studies on barriers to VR application testing and formulate techniques to overcome them. Moreover, \texttt{RQ3} clearly identifies some of the test smells in VR test code. These smell categories are 
%  primarily 
 derived from traditional software test smells.
%  application test-quality issues. 
 Since VR applications are different from traditional software, researchers can investigate the existence of VR-specific test smells such as rendering smells, GameObject configuration smells, etc. Finally, through \texttt{RQ4}, we identified some common categories of test cases developed for VR applications. Such categorization can be a basis for future research on pattern-based automatic test case generation for VR applications.
% \vspace{-0.2cm}
\section{Conclusion}
\vspace{-0.15cm}
\label{sec:conclusion}
%%\todo{Xue}
In this paper, we conducted the first quantitative and qualitative study on existing test practices of VR applications. 
We developed the first tool for  test effectiveness analysis and test-smell identification
% support
for Unity-based projects. 
Moreover, we manually explored the characteristics of VR test cases and categorized them.
Our automatic analysis shows a low adoption of testing by VR applications, and that their test practices are less efficient and have a lower design quality.
% The test-to-code ratio remains lower than comparable projects on the scale of methods, and classes. Their average code coverage and assertion rates are lower than traditional software with value of 15.63\% and 5.96\%.
% Finally, 4  test smells exhibit rates vary from an average of 38.43\% to 3.01\% of a projects' tests.
Furthermore, our manual analysis
% of 220 VR test cases 
resulted in a VR test taxonomy composed of ten main categories,
% such as physics testing, animation testing, etc., 
which reflect the characteristics and specificities of VR applications, along with the identification of VR-specific test categories.
% and our analysis of 281 non-VR tests
% helped us extract which test categories are specific to VR.
We hope that our findings on testing practices in VR applications will allow future researchers to determine VR testing challenges and inspire future research on VR test automation.
% For future works, We plan to test the compatibility and effectiveness of automatic test generation techniques in testing VR applications.
\newpage
\bibliography{main}

\end{document}